# "Close...but not as good as an educator" - Using ChatGPT to provide formative feedback in large-class collaborative learning

**Cory Dal Ponte, Sathana Dushyanthen, Kayley Lyons**
Centre for Digital Transformation of Health, University of Melbourne
cory.dalponte@unimelb.edu.au

**ABSTRACT**: Delivering personalised, formative feedback to multiple problem-based learning groups in a short time period can be almost impossible. We employed ChatGPT to provide personalised formative feedback in a one-hour Zoom break-out room activity that taught practicing health professionals how to formulate evaluation plans for digital health initiatives. Learners completed an evaluation survey that included Likert scales and open-ended questions that were analysed. Half of the 44 survey respondents had never used ChatGPT before. Overall, respondents found the feedback favourable, described a wide range of group dynamics, and had adaptive responses to the feedback, yet only three groups used the feedback loop to improve their evaluation plans. Future educators can learn from our experience including engineering prompts, providing instructions on how to use ChatGPT, and scaffolding optimal group interactions with ChatGPT. Future researchers should explore the influence of ChatGPT on group dynamics and derive design principles for the use of ChatGPT in collaborative learning.

## 1  INTRODUCTION

In blended learning, collaborative problem-based learning frequently serves as a method for having students apply their understanding acquired in self-directed learning modules (Hmelo-Silver & DeSimone, 2013). Ideally, each problem-based learning group would have a facilitator supporting and providing valuable formative feedback, checking for misunderstandings, and encouraging discussion. However, similar to many courses, in our Applied Learning Health Systems professional development course, we only have two instructors for the ten groups completing activities in virtual break-out rooms. Class-wide feedback is provided in debriefs after the activity; however, this feedback does not provide specific feedback for each group.

To solve our problem, we used ChatGPT to provide groups with personalised formative feedback that supported their learning within a large-scale collaborative learning virtual workshop. The tool provided a new feedback loop where groups could discuss and respond to immediate personalised feedback from ChatGPT (see Figure 1).

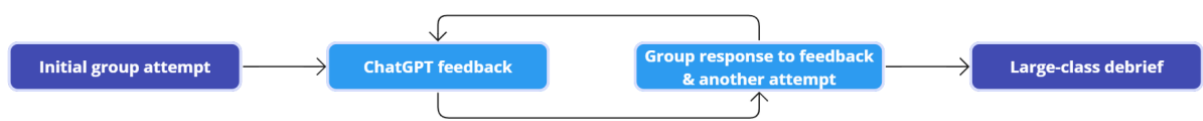

**Figure 1. Use of ChatGPT in the activity provides a new feedback loop**

To evaluate the potential benefits and unintended consequences of the new activity design, we posed the following research questions:
1. How did participants perceive the quality of the feedback provided by ChatGPT?
2. How did participants describe the impact on their learning and group dynamics?
3. What reasons do students attribute to the quality ratings (e.g., Beginner, Advanced) received from ChatGPT on their evaluation plans?

## 2 METHODS

We used the computer-supported collaborative learning (CSCL) design framework proposed by Zheng (2021) to describe our activity and integrate ChatGPT for formative feedback. The course was a 13-week professional development course for practicing health professionals. Each week a different digital health topic (e.g., implementing digital solutions) is presented based on the Learning Health Systems framework. The activity was piloted during one week of the course. The goal for this week was to enable participants with no prior research training to craft a concise evaluation plan that incorporated fundamental components and ensured alignment with the research question. Before the workshop, participants completed three hours of self-directed modules. Then, in the 2.5-hour workshop, the ChatGPT activity was completed in a 45-minute Zoom® breakout room activity with groups of 5-7 participants. Prior to the activity, we provided a 15-minute ChatGPT demonstration and detailed step-by-step instructions. Groups had 25 minutes to craft their evaluation plan followed by 20 minutes for participants to gather and react to ChatGPT feedback.

We've included a link to view the conversation with GPT4 to engineer the instruction and prompts. As ChatGPT requires structured instructions to get the desired output (Jowsey et al., 2023), we engineered the following set of standardized custom instructions and prompts for students to use within the activity (see Figure 2):

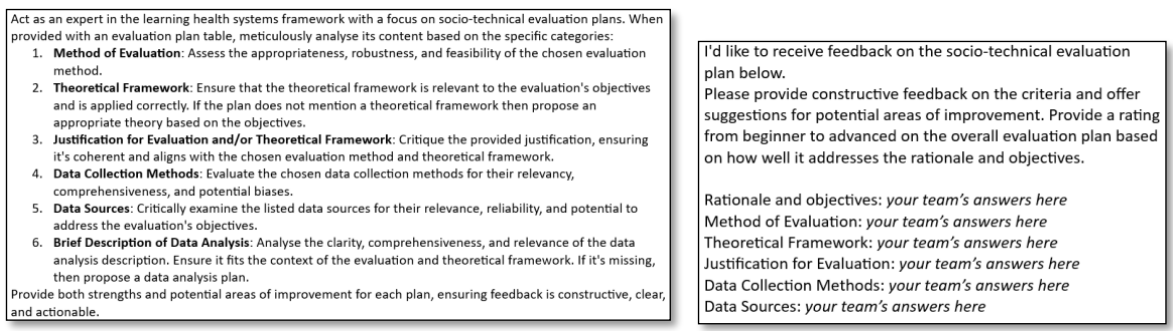

Figure 2. ChatGPT Custom Instructions and Prompts

This study was approved by the University of Melbourne human ethics committee (Project ID 22641). Following the activity, participants immediately completed a Qualtrics® survey. The survey included a mix of open-ended questions and Likert scales according to the research questions. Likert scales were presented as descriptive statistics. Open-ended questions were analysed by a qualitative researcher who open-coded responses according to the research questions. After themes were determined by open-coded, we conducted a frequency analysis of themes.

## 3 RESULTS

The session included 55 participants from 13 organisations (92% hospitals, 8% University) and 18 job titles (45% health professions, 18% health services management, 13% researchers, 18% IT/data analytics, 5% consumers). 44 participants (80%) completed the survey. Half of the participants (51%, n=23) had never used ChatGPT before, with 27% (n=12) using it once a month, 16% (n=7) weekly and 5% (n=2) daily. Through an analysis of the groups' conversation record with ChatGPT, we identified that the majority (7 out of 10) of groups did not iterate their evaluation plans after receiving one round of feedback. We asked participants how they perceived the quality of feedback they received (see Table 1.)

Table 1: Perceived quality of feedback received from ChatGPT (n=44)

| Field | Strongly agree | Agree | Neither agree nor disagree | Disagree | Strongly disagree |
|---|---|---|---|---|---|
| The ChatGPT feedback was valuable | 32% | 52% | 11% | 5% | 0% |
| The ChatGPT feedback helped me learn about digital health evaluation | 16% | 52% | 20% | 11% | 0% |
| The ChatGPT feedback influenced us to improve our evaluation plan | 30% | 55% | 14% | 2% | 0% |
| The ChatGPT feedback increased my interest in digital health evaluation | 16% | 32% | 36% | 16% | 0% |
| The ChatGPT feedback increased my interest in generative AI | 45 % | 41% | 11% | 2% | 0% |

*Influence on learning*: In response to the question "How did the personalized feedback from ChatGPT influence your learning about digital health evaluations? If anything?," most participants wrote that ChatGPT provided specific (n=12), actionable (n=9), immediate (n=7), and easy to understand (n=3) feedback on their evaluation plans. Six participants wanted the feedback validated by a human before deciding to learn from the feedback, or noted they did not have the expertise to judge the quality of the output. Exemplifying the mixed comment group was one participant who noted that the feedback was "close but not as good as an educator." Negative written comments (n=6) either simply stated ChatGPT did not influence their learning, or the feedback was too vague.

*Influence on group dynamics*: We found a wide variety of written responses regarding how ChatGPT influenced group dynamics. Twenty-four participants noted how it enhanced group discussion due to the group responding to the feedback (n=16), becoming behaviourally engaged due to the novelty of ChatGPT (n=4), and ChatGPT activating the discussion (n=3). Six participants wrote that ChatGPT did not influence group dynamics. Surprisingly, 14 participants described how ChatGPT hindered their group dynamics either by providing a distraction, causing group members to spend much more time reading the output, or providing an authority within the group. For example, one participant noted, "ChatGPT essentially acted as the final arbiter." Another participant described that it hindered discussion because "with ChatGPT in the room, there was less need for general discussion as we could simply ask it for answers."

*Causal attributions*: In response to the survey question, "Why do you think your group received either a beginner, intermediate, or advanced rating from ChatGPT?," rarely did participants not trust or disagree with their rating. In terms of causal attribution theory (Kelley, 1973), 31 participants gave descriptions that were coded as adaptive responses to feedback (i.e., internal, non-stable, in-control cause, e.g., "we didn't give enough information on what the data analysis part looked like in-depth"). Counter to our expectations, only one participant distrusted their GPT rating. This participant said their rating was received "because it [ChatGPT] does not know enough."

## 4    DISCUSSION & RECOMMENDATIONS

Overall, we demonstrated favourable perceptions from a group of learners who were mostly new to ChatGPT and practicing professionals. ChatGPT facilitated an additional feedback loop within the learning activity, a feat that would have been unattainable without the deployment of ten standby instructors. Part of the increased engagement and interest in the activity could be explained by the novelty effect of new technology although a few students mentioned this as a distraction.

In future iterations of this activity, we aim to improve the specificity of the ChatGPT feedback through further prompt engineering. As many groups' answers were lacking in detail due to the time constraints, we will increase the focus of prompts on suggesting examples on how to improve their inputs. Also, we will enhance the structure of the activity to ensure the learners return to their evaluation plans and improve them.

We used OpenAI ChatGPT 3.5 interface with equity and scalability in mind. Future educators can learn from the design of our activity and prompts. Based on our findings, for educators integrating generative AI (genAI) into large class size collaborative learning environments, we recommend:
1. Engineer and rigorously test standard custom instructions and prompts
2. Provide detailed instructions outlining how students should interact with the tool
3. Scaffold how the group should optimally interact with ChatGPT

Future research should focus on identifying sets of design principles that assist educators to optimally utilise genAI within their contexts. Given the diversity in group dynamics in our study, future researchers should observe and investigate the factors for why some groups had rich discussions while others had superficial or no discussion at all. Also, researchers should continue to explore student beliefs with using genAI.

## REFERENCES

Hmelo-Silver, C.E. & DeSimone, C. (2013). Problem-based learning: An instructional model of collaborative learning. In *The international handbook of collaborative learning*. Routledge, pp. 370-385.

Jowsey, T., Stokes-Parish, J., Singleton, R., & Todorovic, M. (2023). Medical education empowered by generative artificial intelligence large language models. *Trends in Molecular Medicine*. https://doi.org/10.1016/j.molmed.2023.08.012

Kelley, H. H. (1973). The processes of causal attribution. *American psychologist*, *28*(2), 107.

Zheng, L. (2021). An Innovative Framework for Designing Computer-Supported Collaborative Learning. In: Data-Driven Design for Computer-Supported Collaborative Learning. Lecture Notes in Educational Technology. Springer, Singapore. https://doi.org/10.1007/978-981-16-1718-8_1